\documentclass[]{pasj02} 
\usepackage[switch,mathlines]{lineno} 
\usepackage{graphicx,physics,comment, siunitx, natbib}
\usepackage{appendix}
\usepackage{makecell} 
\jyear{2024}
\Received{}
\Accepted{}

\begin{document} 

\title{Optical photon-counting observation of the Crab pulsar with Kanata telescope using prototype IMONY }

\author{
Takeshi \textsc{Nakamori}\altaffilmark{1}$^{,\dag}$,\orcid{0000-0002-7308-2356}\email{nakamori@sci.kj.yamagata-u.ac.jp}
        Kazuaki \textsc{Hashiyama}\altaffilmark{1,2}$^{,\dag}$\email{khashi@icrr.u-tokyo.ac.jp}
        Rin \textsc{Sato}\altaffilmark{1},
        Masayoshi \textsc{Shoji}\altaffilmark{3},
        Anju \textsc{Sato}\altaffilmark{1},
        Eiji \textsc{Ono}\altaffilmark{1},
		 Yuga \textsc{Ouchi}\altaffilmark{1},
		 Tatsuya \textsc{Nakaoka}\altaffilmark{4},
         Koji \textsc{S.~Kawabata}\altaffilmark{4,5},
		 Toshio \textsc{Terasawa}\altaffilmark{6},
		 Hiroaki \textsc{Misawa}\altaffilmark{7},
		 Fuminori \textsc{Tsuchiya}\altaffilmark{7},
         Kazuhiro \textsc{Takefuji}\altaffilmark{8},
         Yasuhiro \textsc{Murata}\altaffilmark{9},
		 Dai \textsc{Takei}\altaffilmark{10,11} \orcid{0000-0001-7415-3446},
 		 Kazuki \textsc{Ueno}\altaffilmark{12},
		 Hiroshi \textsc{Akitaya}\altaffilmark{13,4}\orcid{0000-0001-6156-238X}
}
\altaffiltext{1}{Faculty of Science, Yamagata University, 1-4-12 Kojirakawa, Yamagata 990-8560, Japan }
\altaffiltext{2}{School of Science, The University of Tokyo, 7-3-1 Hongo, Bunkyo, Tokyo 113-8654, Japan}
\altaffiltext{3}{KEK, High Energy Accelerator Research Organization, Institute of Particle and Nuclear Studies, 1-1 Oho, Tsukuba, Ibaraki 305-0801, Japan}
\altaffiltext{4}{Hiroshima Astrophysical Science Center, Hiroshima University, 1-3-1 Kagamiyama, Higashi-Hiroshima, Hiroshima 739-8526, Japan}
\altaffiltext{5}{Graduate School of Advanced Science and Engineering, Hiroshima University, 1-3-1 Kagamiyama, Higashi-Hiroshima, 739-8526, Hiroshima, Japan}
\altaffiltext{6}{Institute of Cosmic-Ray Research, University of Tokyo, 5-1-5 Kashiwanoha, Kashiwa, 277-8582 }
\altaffiltext{7}{Planetary Plasma and Atmospheric Research Center, Tohoku University, 6-3 Aramaki Aza Aoba, Aoba, Sendai, 980-8578, Japan
}
\altaffiltext{8}{Usuda Deep Space Center, Japan Aerospace Exploration Agency, 1831-6 Omagari, Kamiodagiri, Saku, Nagano 384-0306, Japan}
\altaffiltext{9}{Faculty of Engineering, Fukui University of Technology, 3-6-1 Gakuen, Fukui, 910-8505, Japan}
\altaffiltext{10}{Daiphys Technologies LLC, 1-5-6 Kudan-Minami, Chiyoda, Tokyo 102-0074, Japan }
\altaffiltext{11}{Research Center for Measurement in Advanced Science, Rikkyo University, 3-34-1 Nishi-Ikebukuro, Toshima, Tokyo 171-8501, Japan}
\altaffiltext{12}{Graduate School of Science, Osaka University, 1-1 Machikaneyama, Toyonaka, Osaka,  560-0043, Japan}
\altaffiltext{13}{Astronomy Research Center, Chiba Institute of Technology, 2-17-1 Tsudanuma, Narashino, Chiba 275-0016, Japan}



\maketitle

\begin{abstract}
We have developed an optical photon-counting imaging system, IMONY, 
as an instrument for short-scale time-domain astronomy. 
In this study, we utilized a Geiger avalanche photodiode array with a $4\times 4$ pixel configuration, 
with each pixel measuring \SI{100}{\micro m}. 
We a a dedicated analog frontend board 
and constructed a data acquisition system with an FPGA 
to time-stamp each photon with a time resolution of \SI{100}{\ns}. 
We mounted a prototype model of the system on the 1.5-m Kanata telescope,
intending to observe the Crab pulsar 
and conduct joint observations with Iitate and Usuda radio telescopes in Japan. 
We successfully demonstrated that IMONY could image the Crab pulsar as an expected point source
and acquire the well-known pulse shape. 
We found that the time lag between the optical and radio main pulses was $304\pm$\SI{35}{\mu s}, 
consistent with previous studies.
\end{abstract}


\section{Introduction}

Time-domain astronomy has significantly progressed in understanding time-varying or transient astronoamical phenomena in recent years. 
Optical and infrared astronomy has encountered difficulty with short timescales, 
but this has been overcome with the development of detectors. 
For instance, high-cadence survey observations with large-area Complimentary Metal Oxide Sensor (CMOS) cameras 
have been successfully implemented \citep[e.g.,][]{sako18,Huang+21}. 
Basically the timescale of a phenomenon is inversely proportional 
to the size of its emission region. 
Observations on very short timescales, milliseconds or less, 
are primarily used to probe phenomena in the vicinity of compact objects. 
Achieving high temporal resolution with CMOS is a challenging endeavor. 

In such contexts, photon counting methods are particularly effective.
The use of single-photon avalanche diode (SPAD) photon-counting detectors in several previous projects,
such as OPTIMA \citep{ste+08}, SiFAP \citep{amb+14}, 
SiFAP2 \citep{Ghe18}, SiFAP4XP \citep{Ghe22}, Iqueye \citep{Nal09}, and Aqueye+ \citep{zam+15}, 
has enabled the acquisition of optical observations with high time resolution up to sub-$\mu$s.
Another example of a sophisticated detector, ARCONS, has been developed by \citet{maz+13},
an optical to near-infrared spectrophotometer 
based on the microwave kinetic inductance technique.
They have reported many scientific results, such as the detection of optical pulsation from a millisecond pulsar \citep{amb+17,zam+19,amb+21}, 
optical enhancement corresponding to the giant radio pulses (GRPs) of the Crab pulsar \citep{she+03,str+13}.
These efforts and demonstrations of performance activated the field of ultra-fast optical astronomy.

We also started developing an optical photon-counting imager 
that consists of a Geiger avalanche photodiode (GAPD, a.k.a. SPAD) array
that is made of a customized Multi-Pixel Photon Counter (MPPC).
Following the successful first demonstration 
of the Crab pulsar detection with a 35-cm telescope \citep{naka+21}, 
we have triggered a project, 
\textit{Imager of MPPC-based Optical photoN Counter from Yamagata} (IMONY),
and upgraded the readout and data acquisition system as a prototype.
A distinctive feature of our instrument is the capability 
to realize imaging with monolithic GAPD arrays. 
The imaging system can have a higher light collection efficiency for point sources than single-pixel systems,
as a viewpoint of the containment of point spread function.
Light collection efficiency is essential 
because photon statistics is often crucial 
in highly time-resolved observations of fast phenomena. 

Another advantage of imaging observation is the stability of photometry.
Even for point sources, the position of the image on the focal plane is not stable due to fluctuations caused by the atmosphere and the telescope's tracking.
The image data is helpful to confirm whether the whole spot is contained in the sensor area. 

ARCONS \citep{maz+13} also achieves multi-pixel imaging, 
even with spectroscopy capability,
but requires cryogenic cooling and large mechanical structures.  
In contrast, our system can be operated at room temperature 
and has the advantage of being compact and easy to handle.

This paper reports on the newly developed readout system
and Crab pulsar observation with the Kanata telescope in coordination with radio telescopes as performance tests.

\section{System}

\subsection{Overview}

\begin{figure}
    \centering
    \includegraphics[width=.9\linewidth]{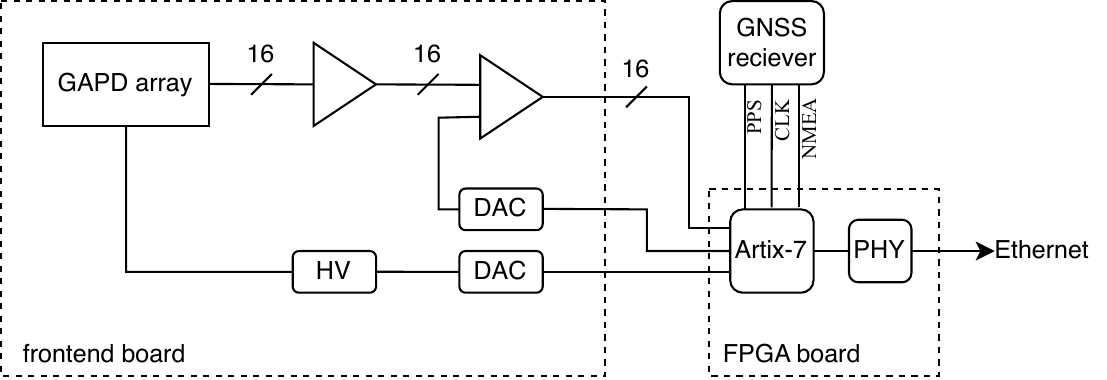}
    \caption{Overview schematics of the system. See text for details.
    {Alt text: Block diagram describing the signal flows and connections from the frontend board to the FPGA board.}}
    \label{fig:system}
\end{figure}

\begin{figure}
    \centering
    \includegraphics[width=.9\linewidth]{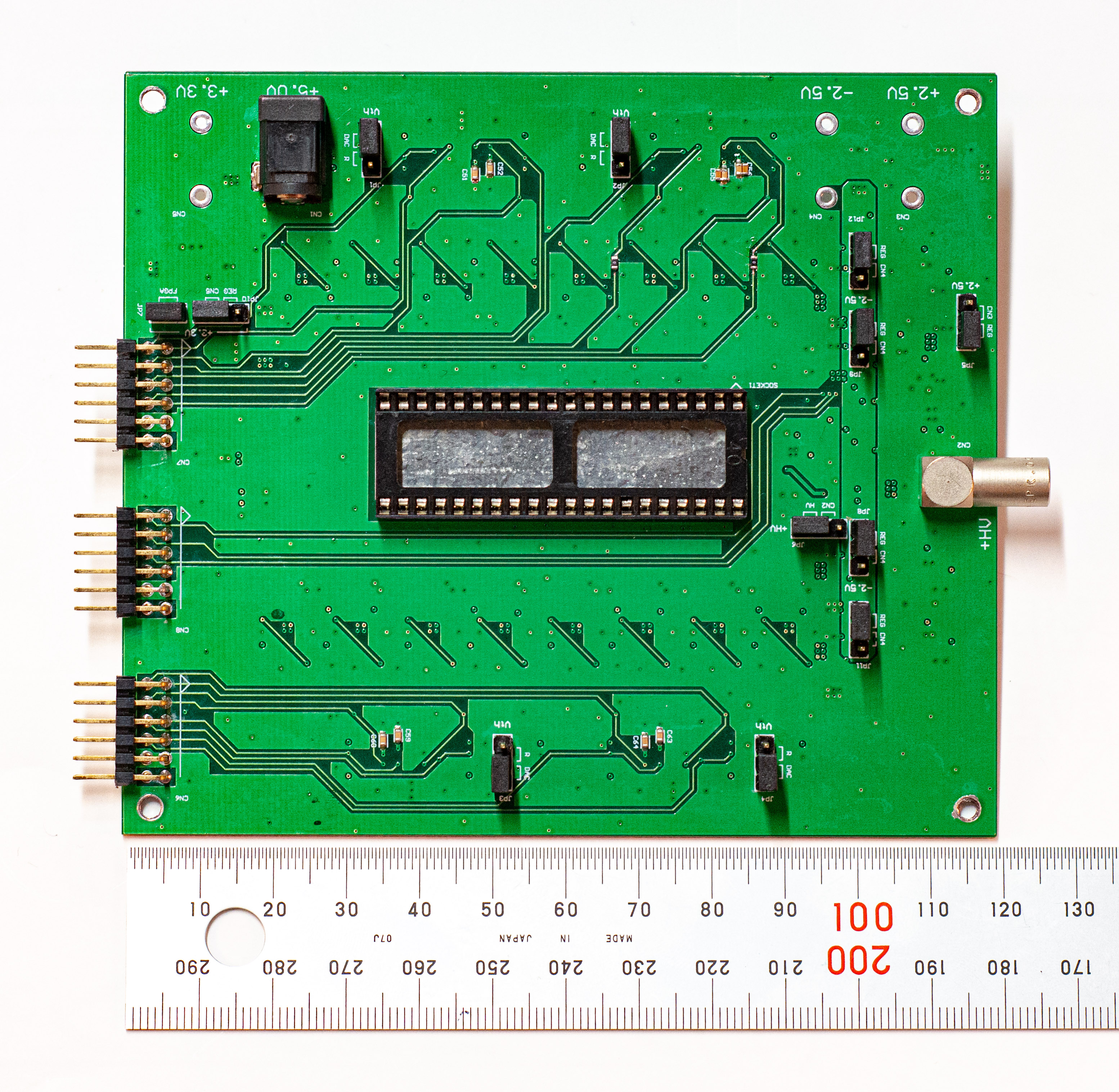}
    \caption{A photo of the frontend readout board. 
    {Alt text: Photo of the frontend board with a ruler indicating the board size as 12 times 12 square centimeters.}}
    \label{fig:tako}
\end{figure}

Figure\,\ref{fig:system} shows an overview of the acquisition system updated in this work. 
The system consists of roughly two parts: a frontend board (FEB), 
where the GAPD array sensor is mounted along with amplifiers and comparators, 
and a data processing board using a Field Programmable Gate Array (FPGA). 
The FEB was designed to make the system volume smaller than the initial version \citep{naka+21} 
that contained power and data bus crates, 
and a commercial FPGA evaluation board was used to reduce cost. 
Figure\,\ref{fig:tako} is a photo of the FEB where a ruler indicates the board size.
The sensor is mounted directly on this board, and each sensor channel's output signal is connected to a wideband amplifier.
Each signal is then fed into a comparator 
to generate a timing signal for photon detection at each pixel. 
A total of 16 lines are sent in parallel to the FPGA board.
A Serial Peripheral Interface (SPI) communication signal from the FPGA also controls the digital-to-analog converter (DAC) 
to set the comparator threshold. 
The SPI controls another DAC and the high-voltage output. 
The following FPGA board completely powers the board.

The FPGA evaluation board is an Artix-7 model that operates with a \SI{10}{MHz} clock 
provided by the Global Navigation Satellite System (GNSS) module as the system clock. 
The GNSS module also provides pulse-per-second (PPS) pulses 
and the National Marine Electronics Association (NMEA) sentences 
to the FPGA via Universal Asynchronous Receiver Transmitter (UART) communication.

We implemented two acquisition modes. 
The first, the \textit{light-curve mode}, is the main mode of observation, 
where we record the arrival time of a photon per pixel as an event list.
In the FPGA, a counter is incremented by a \SI{10}{\MHz} clock 
and is reset each time when a PPS is input. 
Another counter runs simultaneously for the PPS 
and is initialized to zero at the start of the measurement. 
When the hit signal is delivered, 
the value of these two counters determines the relative time 
at which the photon was detected with an accuracy of \SI{100}{\ns}.
Combined with this relative time stamp, the UTC obtained from the NMEA sentences at the measurement's beginning 
yields the absolute photon detection time.
Although a \SI{5}{\ns}-clock is used inside the FPGA 
to determine the timing of a hit signal, 
a timestamp with a resolution of \SI{100}{\ns} is given in this version, 
considering expected photon flux and reducing the data size.
The dead time caused by the readout is negligible due to the presence of a FIFO buffer. 
However, each cell becomes insensitive for approximately 100 ns after a hit during its recovery time.

The second measurement mode, the \textit{scaler mode}, 
counts the number of pulses detected during a specified exposure for each channel, 
namely a count map, where the data is stored in the internal registers.
Since the data size is drastically reduced and easier to handle,
we implemented this mode for quick health checks during start-up procedures or observations.

The data transfer and command interfaces are implemented using SiTCP \citep{uchida08}, 
which has a high-speed transfer capability using TCP/IP.
This work does not realize the full transfer rate 
since our evaluation board only supports 100BASE-T. 
As described later, however,
it is sufficient for the data transfer rate required for observation.
SiTPC can also handle slow control using User Datagram Protocol (UDP) over the ethernet
and enable access to the internal bus and resisters, the so-called Remote Bus Control Protocol (RBCP).
RBCP is employed to retrieve data acquisition and the DAC control 
and to read out the data stored in the resisters, 
such as NMEA information from the GNSS and count numbers obtained by the \textit{scaler-mode}.

\subsection{Evaluation}

\begin{figure}
    \centering
    \includegraphics[width=\linewidth]{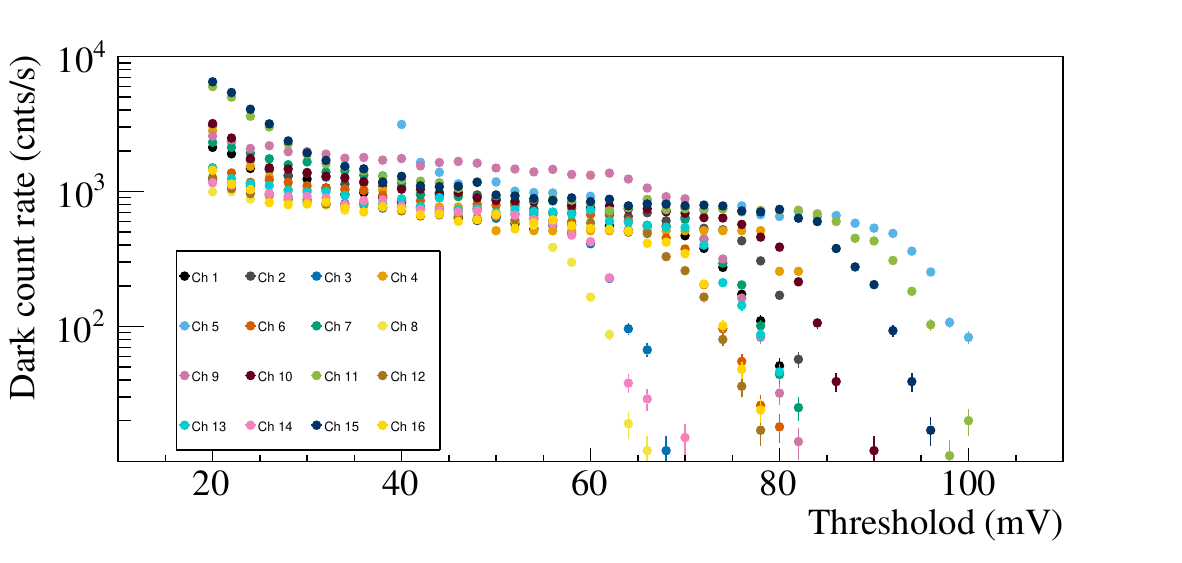}
    \caption{Example of \textit{dark-scan} result taken at the temperature of 0${}^\circ$C.
    {Alt text: A set of 16 plots. The horizontal axis range 10 to 120 mV, while the vertical axis spans 10 to 10 to the forth counts per second.}}
    \label{fig:darkscan}
\end{figure}

We performed calibration and performance tests in the laboratory,
then the signal delay and timing jitter were evaluated.
After the calibration of the DACs for the high voltage and threshold settings,
we implemented the \textit{dark-scan} functionality, 
where we simultaneously measured the dark count rate for all 16 pixels
for various threshold values.
We obtained a set of \textit{dark-scan} curves, an example of which is shown in Figure,\ref{fig:darkscan}.
Since the intrinsic gain variation among the GAPD cells is 3\% \citep{naka+21},
the apparent differences in the cutoff at higher threshold voltages are 
attributed to the gain and offset variations 
caused by the characteristics of the amplifier on the FEB.
It is also worth noting that channel 9 is known to exhibit a higher dark count rate 
compared to the others \citep{naka+21}.
Finally, we decide the threshold value for the operation 
by finding a common value on the plateau of the \textit{dark-scan} curves 
since the current system version can only set the common threshold voltage by the single DAC chip.

Using an oscilloscope, we observed a signal delay of 30--\SI{40}{\ns} before and after the comparator I/O. 
On the other hand, jittering was negligible compared to the delay time scale.
Then we irradiated laser pulses of \SI{80}{\ps} wide \citep{inome+19}, 
far shorter than the time resolution of IMONY, 
to the GAPD array through an appropriate Neutral Density filter.
This pulsed laser was triggered by a PPS pulse from another GNSS receiver, which was separate from the system. 
The time-of-arrival (TOA) of the laser beam was measured, 
and the variation in timing indicated 
that more than 90\% of events occurred within a single time bin of 100 ns, corresponding to the PPS. 
The remaining events appeared in the subsequent time bin.
This result is consistent with the overall timing accuracy of the PPS, 
the timing jitter of the laser trigger detection, 
and the delay of the FEB. 
Therefore, we conclude that, in the most conservative estimation, 
IMONY has an absolute time accuracy of less than \SI{200}{\ns}.

\section{Observation}

\subsection{Radio}

\begin{longtable}{cccccr}
  \caption{Radio observation summary }\label{tab:radioobs}  
\hline\noalign{\vskip3pt} 
		Observatory & $\nu$ &$\Delta \nu$  & Resolution &  Start time & Duration  \\ 
        & (MHz) & (MHz) & (bits/sample) & (UT) & (hour)\\
        \hline\noalign{\vskip3pt} 
\endfirsthead      
\hline\noalign{\vskip3pt} 
  Name & Value1 & Value2 & Value3 \\  [2pt] 
\hline\noalign{\vskip3pt} 
\endhead
\hline\noalign{\vskip3pt} 
\endfoot
\hline\noalign{\vskip3pt} 
\multicolumn{2}{@{}l@{}}{\hbox to0pt{\parbox{160mm}{\footnotesize
\hangindent6pt\noindent
}\hss}} 
\endlastfoot 
 	Iitate	&  325.1  & 16.0 & 2 & 2021 Dec 5 10:00:01 & 10.17 \\
  		Usuda &  2256.0  & 128.0 &4 & 2021 Dec 5 15:50:00 & 2.67 \\
		  Usuda  & 8438.0  & 128.0 &4 & 2021 Dec 5 15:50:00 & 2.67 \\
            Iitate & 325.1& 16.0 &2 & 2021 Dec 6  09:57:01 & 10.64 \\
	    Iitate & 325.1&16.0 &2 & 2021 Dec 7 10:00:01 & 10.20 \\
            Iitate& 325.1 & 16.0 &2& 2021 Dec 8  09:49:01 & 10.64\\ 
\end{longtable}

To obtain the radio timing parameters of the Crab pulsar,
which are needed for optical timing analysis,
we performed joint observations with the Iitate (Fukushima, Japan) \citep{tsuchiya+10} 
and Usuda (Nagano, Japan) radio telescopes.
A summary of observations is shown in Table \ref{tab:radioobs}.
Iitate Observatory covers the P-band with Iitate Planetary Radio Telescope, 
which consists of dual asymmetric offset parabolas with dimensions of $31 \times 16.5$\,m$^2$ for each.
Data were acquired in a single channel with a resolution of 2 bits and a bandwidth of 16 MHz.
Usuda Observatory is responsible for the S- and X-bands, employing a single dish parabola with a diameter of \SI{64}{\m}.
They recorded data for each band with 4-bit resolution and \SI{16}{\MHz} bandwidth
from four channels, for a total coverage of \SI{128}{\MHz}.
Both telescopes were able to observe on December 5.
However, the Usuda station was not available on December 7 due to snow,
although it was scheduled to observe.
The Iitate station was also available and observed on December 6.

\subsection{Optical}
\label{optobssec}

\begin{figure}
    \centering
    \includegraphics[width=.9\linewidth]{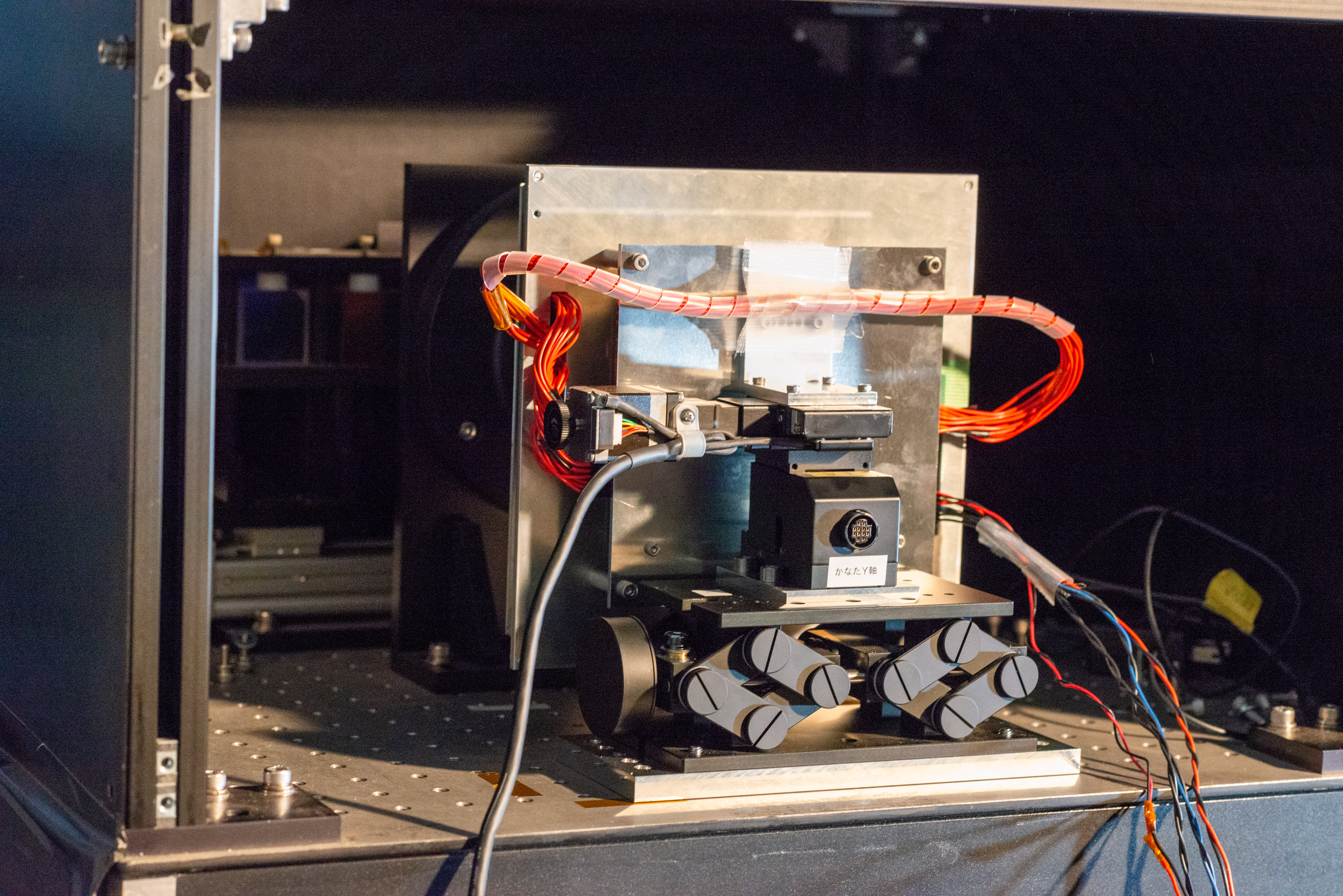}
    \caption{A picture during the installation of the IMONY prototype at the Nasmyth focus of the Kanata telescope. 
    The whole readout system is contained in the aluminum case.
    {Alt text: A picture of IMONY, placed on the optical surface plate.}}
    \label{fig:imo}
\end{figure}

We have made optical observations of the Crab pulsar with the Kanata telescope 
at the Higashi-Hiroshima observatory (Hiroshima, Japan), with an aperture radius of \SI{1.5}{\m} ($F=12.2$).
The location of the telescope is ($132^\circ 46'36"$E, $34^\circ 22'39"$N) at \SI{511.2}{\m} above sea level, 
which we used to calculate TOAs.
Figure \ref{fig:imo} shows the IMONY installation at the Nasmyth focus.
The system was mounted with two motorized stages
for scanning $X$ and $Y$ on the focal plane to compensate for the small field-of-view (FoV) of $\sim 8"$, as performed in the previous study \citep{naka+21}. 
We adjusted the position of the telescope's secondary mirror to focus the image.
The pixel size of $100\,\mu$m corresponds to approximately $1.9"$ on the focal plane. 
We confirmed that a spot size of about 2", 
consistent with the typical seeing for the Kanata telescope, was obtained.

The Crab pulsar was observed from about 14:00 UTC to 19:00 on December 7, 2021.
Data taking was divided into many runs for 1--5 minutes.
The weather was generally clear, although it was sometimes partially cloudy.
We had to repoint the telescope several times during the observation
because the FoV of the sensor was so small 
that the tracking accuracy was not tolerant enough to keep the target in the center of the sensor.
We discuss this aspect later.

As mentioned in the previous section, 
our system is not equipped with any cooling functionality such as a Peltier cooler.
Since the front-end circuit board, especially the amplifiers generates heat, 
the sensor case has a fan to dissipate the warm air. 
The temperature and humidity around the GAPD are monitored 
by Raspberry Pi using a BME280 sensor.
The temperature remained stable around $5{}^\circ$C, 
fluctuating by no more than $\pm 2,{}^\circ$C through the night.

During the observation, we took dark frame data several times by closing the automated mirror facet of the telescope,
and also flat frame data using the built-in flat lamp and panel of the observatory.

\section{Analysis and result}

\subsection{Radio}
\label{sec:radana}

The goal of radio analysis is to obtain the information needed to phase the visible photon time stamps: the rotation frequency and its derivative 
and the reference time of phase zero.
The peak of the main radio pulse is defined as phase zero,
and the TOA of the first pulse on each observation day $t_\mathrm{JPL}$ is determined following the analyses procedures in \citet{enoto+21}.
The timing parameters used in and derived from this work are listed in Table \ref{tab:timing}.

\begin{figure}
    \centering
    \includegraphics[width=\linewidth]{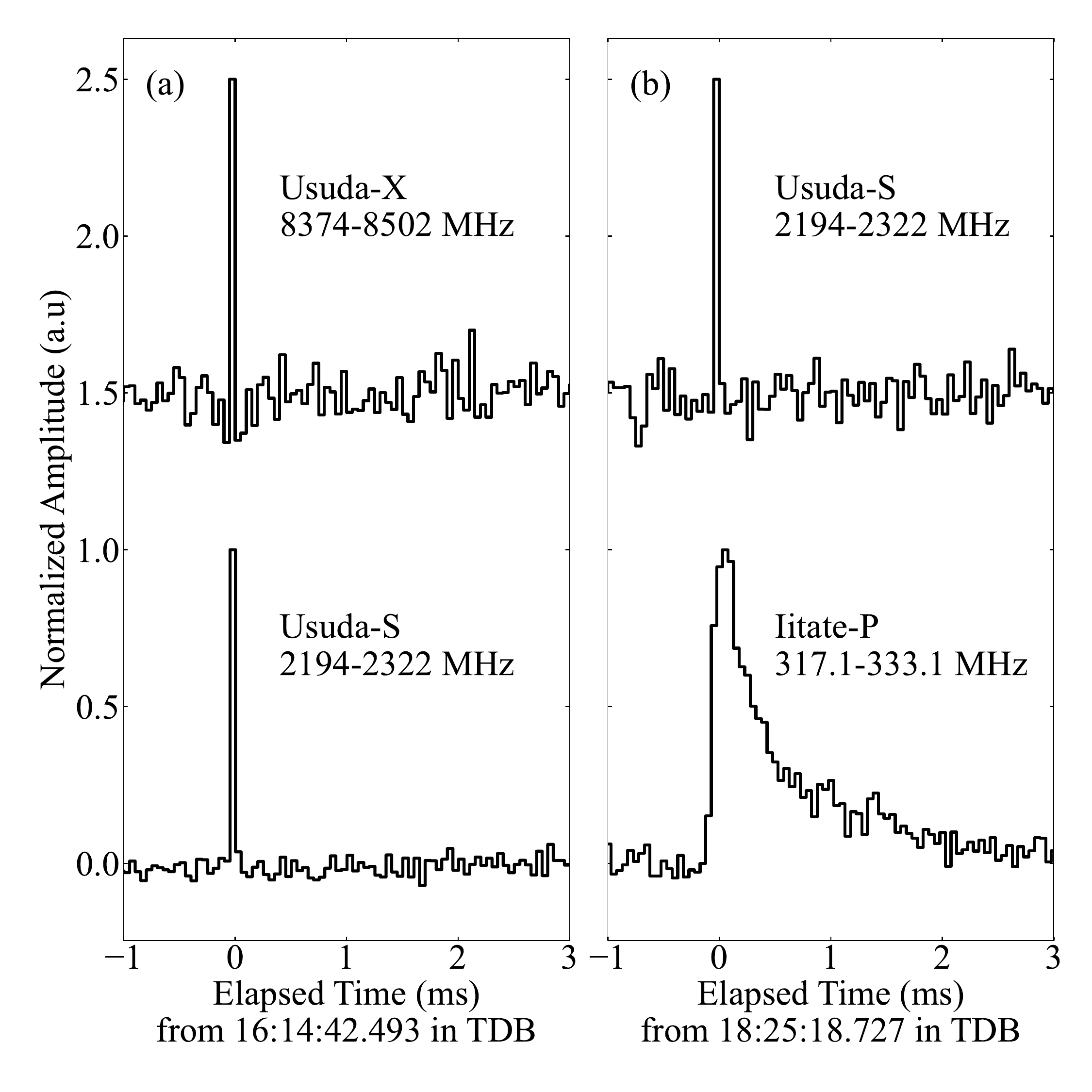}
    \caption{ Examples of GRPs of the Crab pulsar, coincident between the multi-bands.
        Each pulse shape is normalized, and the baselines are shifted for clarity.
    (a) X- (above) and S-band (bottom) coincident events used for estimating the DM.
    (b) X- (above) and P-band (bottom) coincident events used for estimating the lag between the peak time for each.
    {Alt text: Two line graphs presenting the GRP shape.}}
    \label{fig:grps}
\end{figure}

We applied the coherent de-dispersion method \citep{Hankins+75}
to determine the Dispersion Measure (DM) and correct the frequency-dependent group delay during the propagation.
First, we identified several significant GRPs
detected simultaneously in both the S- and X-band Usuda data.
Then, we determined the best DM 
so that the pulses were simultaneous in both bands by correcting for the group delay. 
Figure \ref{fig:grps} shows an example of such a coincident GRP,
while unfortunately, we did not detect the coincident GRP over X-, S-, and P-band.
We did not use the low-frequency P-band to determine the DM 
since the pulse shape shows a non-negligble time lag due to the slow rise.
This effect has been known to be
more significant in lower frequency GHz due to scattering in the interstellar medium (ISM).

\begin{longtable}{llll}
  \caption{Timing parameters of the Crab pulsar used in this work}\label{tab:timing}  
\hline\noalign{\vskip3pt} 
 Parameter & 5 December & 7 December & Source\\  [2pt]
\hline\noalign{\vskip3pt} 
\endfirsthead      
\hline\noalign{\vskip3pt} 
  Name & Value1 \\  [2pt] 
\hline\noalign{\vskip3pt} 
\endhead
\hline\noalign{\vskip3pt} 
\endfoot
\hline\noalign{\vskip3pt} 
\multicolumn{2}{@{}l@{}}{\hbox to0pt{\parbox{160mm}{\footnotesize
\hangindent6pt\noindent
\hbox to6pt{\footnotemark[$*$]\hss}\unskip%
  Jodrel Bank Observatory.
}\hss}} 
\endlastfoot
    Right Ascension & \timeform{5h34m31.97232s} & fixed & JBO\footnotemark[$*$]\\
    Declination & \timeform{22D00'52.069"} & fixed & JBO\\
    Ephemeris & DE430 &  fixed & $-$\\
    Epoch & 59553.00000038667824 & 59555.000000320011573 & this work\\
    First pulse arrival time $t_\mathrm{JPL}$ (ms)& $33.355\pm0.005$ & $27.530\pm0.009$ & this work\\
    Pulsar frequency $\nu$ (s$^{-1}$)& 29.5893614566(2) & 29.5892979477(2) &JBO\\
    Frequency first derivative $\dot{\nu}$ (s$^{-2}$)& $-367527.57(22)\times10^{-15}$ & $-367527.57(22)\times10^{-15}$ & JBO \\
    Frequency second derivative $\ddot{\nu}$ (s$^{-3}$) & $7.3\times10^{-22}$ & fixed & JBO \\
    Dispersion Measure (pc\, cm$^{-3}$) & 56.7420(10) & fixed & this work\\
\end{longtable}

\begin{figure}
	\centering
	\includegraphics[width=\linewidth]{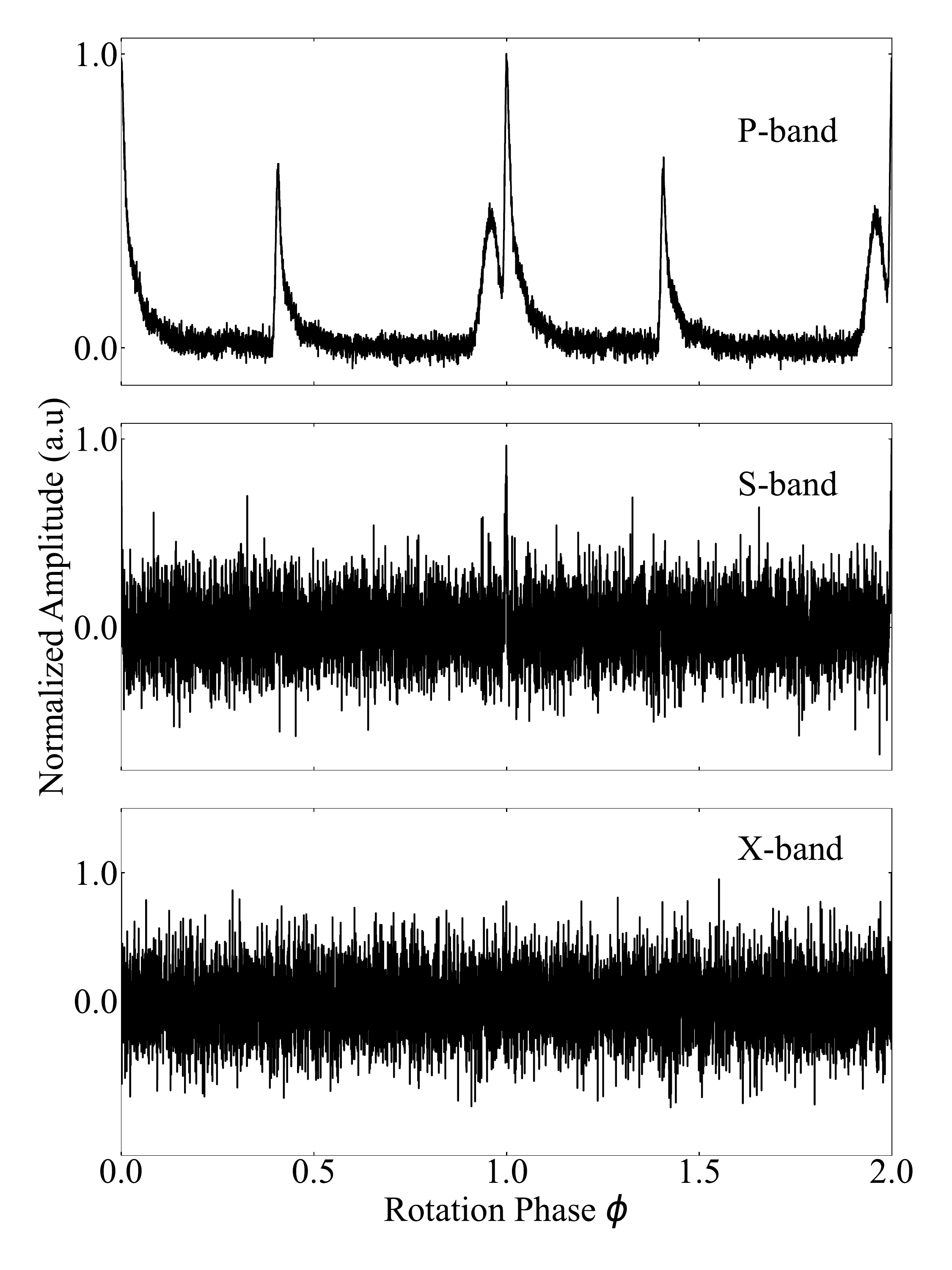}
	\caption{Time-averaged light curves of the Crab pulsar,
            obtained by the Iitate P-band (top), the Usuda S-band (middle), and the Usuda X-band (bottom). 
            Two cycles of the pulses are shown. 
            The integration time bin is \SI{10}{\mu s} for each.
        {Alt text: Three line graphs showing the pulse shapes for the three bands.}}
	\label{fig:avepulse_PSX}
\end{figure}

After the de-dispersion process, 
we transformed the TOAs recorded in UTC into the solar system barycentric dynamical time (TDB)
using the \texttt{PINT} package \citep{Luo+21}.
The timing model used for the transformation is shown in Table \ref{tab:timing}.
Then, we generated time-averaged folded light curves using the pulsar period, its first and second derivatives
calculated by extrapolation of those values provided by the Jodrell Bank observatory \citep{lyne+93}\footnote{http://www.jb.man.ac.uk/~pulsar/crab.html}.
After this procedure we excluded time intervals affected by the radio frequency interference.
As shown in Figure  \ref{fig:avepulse_PSX}, the main pulses were detected in both the P- and S-bands,
while the significant time-averaged signal was not observed in the X-band.
That is, we could detect only the GRPs in this X-band data 
since the time-averaged emission of the Crab pulsar is fainter in higher frequencies.
Finally, we determined $t_\mathrm{JPL}$ to be $33.355\pm 0.005$\,ms for December 5 so that the time-averaged pulse peak of S-band data is aligned best.

Unfortunately, we could not obtain the S-band data on December 7, the day of the optical observation. 
So we must determine $t_\mathrm{JPL}$ only from the P-band data.
It is well known that the intrinsic pulse shape in low frequency was smeared and broadened due to the interstellar plasma, often modeled as scattering by thin screen plasma sheets in the ISM \citep[e.g.,][]{mey07}.
As a result, the peak of the intensity in the P-band arrives later than in the S-band.
We estimated the time lag was $95\pm7\,\mathrm{\mu s}$ in December 5 data.

Variation of the ISM also causes the DM and the decay time of the trailing tail of the GRPs in the P-band.
Long-term monitoring of the DM \citep{mckee18} shows significant variation, 
although the day-scale variation was smoothed out in their studies. 
\citet{mckee18} also reported that the decay time of the pulse, referred as the scattering timescale $\tau _\mathrm{sc}$, is closely correlated with the DM.
We evaluated the decay time of our P-band data day by day,
as an indicator of the ISM density variation.
We found no significant changes in $\tau _\mathrm{sc}$ over the four days of data.
We then assume that the ISM was stable during these days of data.
so that could apply the common DM and time lag on 7 December.
Finally, we estimated $t_\mathrm{JPL}$ as $27.530\pm 0.009$\,ms for December 7 using the P-band data.
The error was derived from the quadrature sum of the errors 
of the integration time (5\,$\mu$s) and the lag between the P- and S-bands (7\,$\mu$s).

\subsection{Optical}

The TOAs of the optical photons in each pixel were reconstructed as absolute UTC times
and then converted to TDB using the barycentric correction function in the \texttt{PINT} package. 
The phase of the Crab pulsar was calculated for each TOA using the radio timing information in Table \ref{tab:timing}, 
which was obtained as described in the section \ref{sec:radana}.
At the same time, the number of rotations of the Crab pulsar 
from $t_\mathrm{JPL}$, called \textit{turns},
were also calculated for each event.

We first generated a light curve in all 16 pixels for each. 
We found the presence of occasional, intermittent rate jumps 
due to electrical noise in the circuits. 
The duration of these noise events was not constant, 
ranging from a few $\mu$s to at most \SI{1}{ms}, 
which was sufficiently shorter than the Crab pulsar rotation period. 
Consequently, a light curve was generated using \textit{turns} as the unit of time,
and all events in the \textit{turns} where the noise was observed were excluded (electric noise cut). 
The cut criteria we applied was $10\sigma$ beyond the averaged rate of the \textit{turn}-wise light curve.
The period during which the count rate decreased or fluctuated significantly
due to the passage of clouds was also cut by run-wise (cloud cut).
The resultant exposure time was \SI{7020}{sec}.

After the data-cleaning procedure by the electric noise and cloud cuts,
we generated folded phasograms using timing parameters listed in Table \ref{tab:timing}
and searched for the Crab pulsation for each pixel on a run-by-run basis.
The detection criterion was $>5\sigma$ excess of the main pulse 
against the off-pulse range, where the on- and off-pulse correspond to $0.990< \phi <1.010$ and $0.7729<\phi <0.8446$ \citep{Slowikowska+09}, respectively.

\begin{figure}
    \centering
    \includegraphics[width=\linewidth]{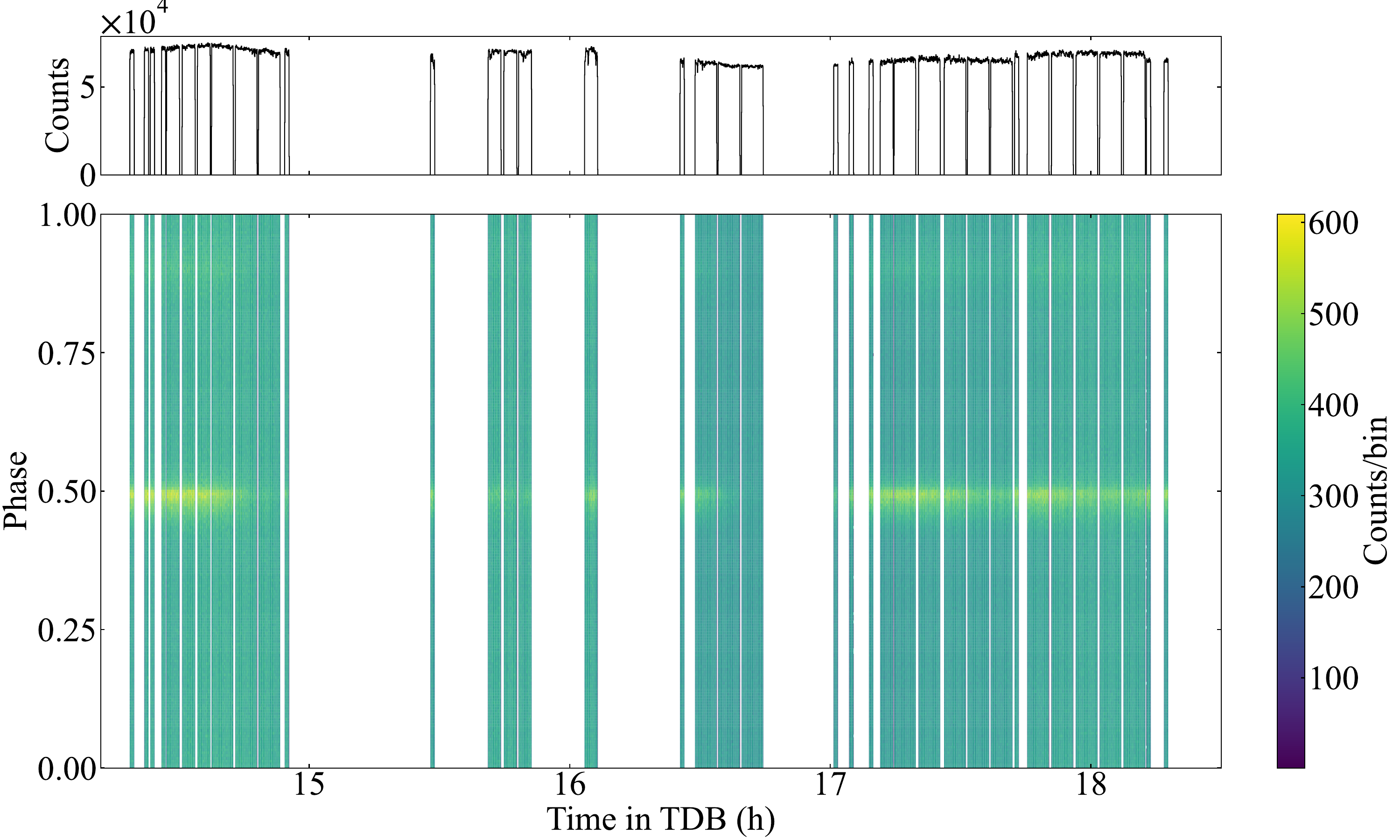}
    \caption{Light curves during the IMONY observation of the Crab pulsar in December 7.
			The horizontal axis indicates the time with a unit of the Crab \textit{turns}.
    			(Top) The count rate summed over all the 16 pixels. 
			(Bottom) The phase distribution of the detected photons along with time. The phase of the vertical axis is shifted by 0.5 to center the main pulse component.
            {Alt text: Two histograms of the optical photon count rates between approximately 14 and 19 hours. The top panel is a 1-D histogram that ranges from 0 to several tens to the forth counts in each bin. In the bottom panel is a 2-D histogram where the phase ranges from 0.0 to 1.0.}}
    \label{fig:opt2d}
\end{figure}

We then applied the dark count subtraction and flat-frame correction.
Figure \ref{fig:opt2d} shows the light curves after the data cleaning.
The count rate in the FoV reflects the sky condition.
As we mentioned in Section \ref{optobssec},
the data taking was split into a few-minute runs,
and the rate drop with a short duration of a few minutes in Figure \ref{fig:opt2d} (top) corresponds to the run switching time.
The bottom panel shows the photon arrival phase distribution over the observing period.
The main peak phase was stable through the night, indicating our system successfully kept working for several hours.
It is also apparent that the count rates around the main peak gradually decrease along with time.
We will discuss this point later.

\begin{figure}
    \centering
    \includegraphics[width=\linewidth]{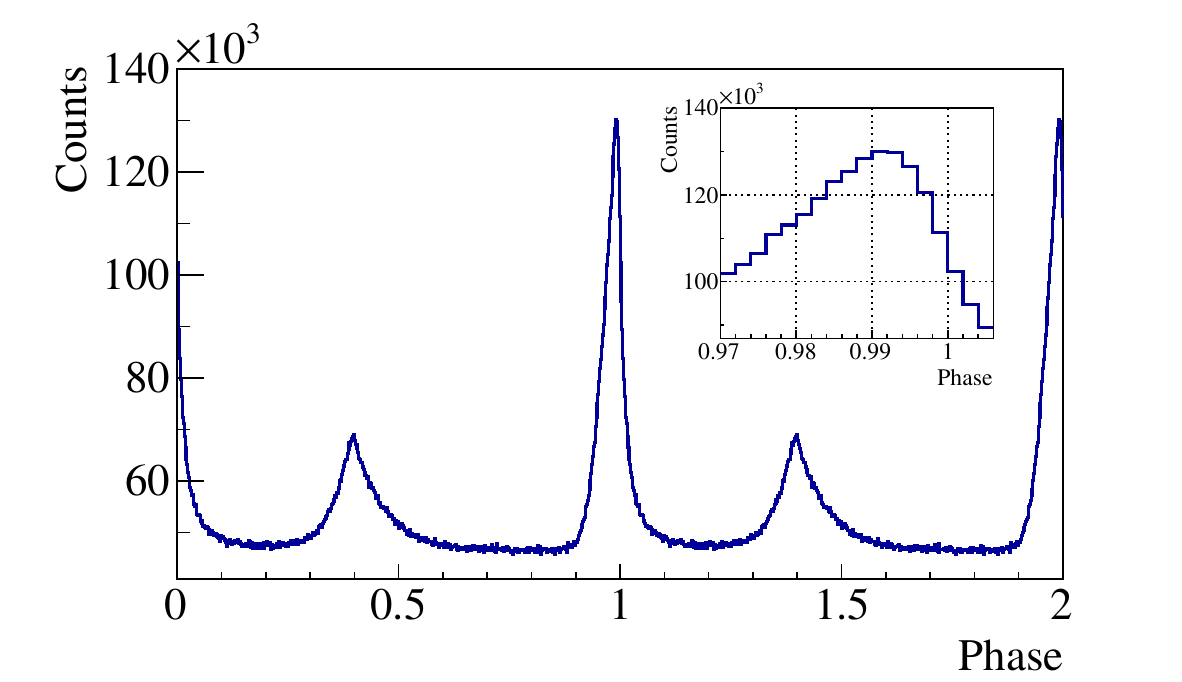}
    \caption{Optical pulse of the Crab pulsar obtained on December 7, 2021.
        The total exposure of 1.95 hours of data is stacked. 
        The optical peak of the main pulse leads to the radio peak 
        by $304\pm35\;\mathrm{\mu s}$.
        {Alt text: Two histograms showing the optical pulse shape. The one is a close-up view drawn in an inset panel.}}
    \label{fig:optresults}
\end{figure}

\begin{figure}
    \centering
    \includegraphics[width=\linewidth]{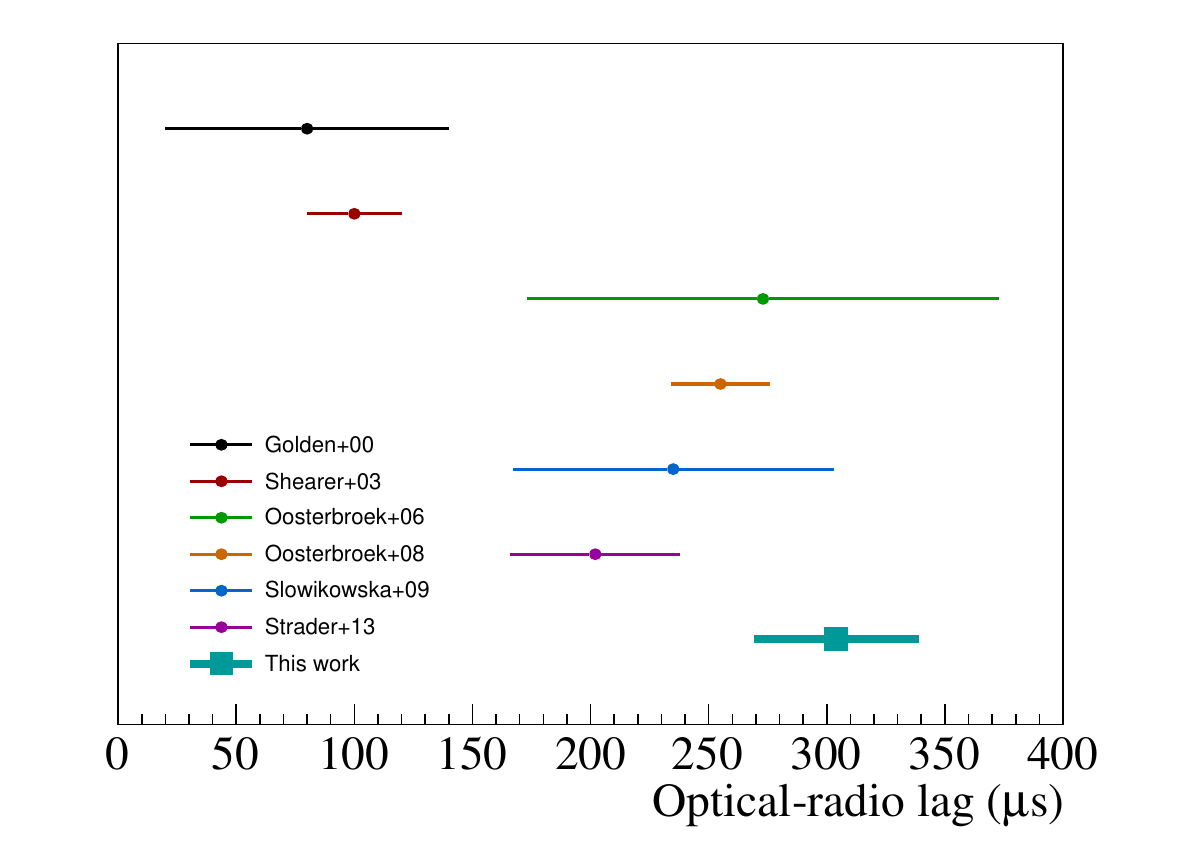}
    \caption{Compilation of previous measurements of the lag between the optical and radio peaks.
            The data were taken from \citet{gol+00}, \citet{she+03}, \citet{oo+06}, \citet{Oosterbroek+08}, \citet{Slowikowska+09} and \citet{str+13}.
        {Alt text: Plots with error bars that indicates the values from each papers.
        The horizontal axis ranges 0 to 300 micro seconds.}}
    \label{fig:lag}
\end{figure}

Figure \ref{fig:optresults} shows a phasogram of the Crab pulsar 
with integrating photon events obtained at pulse-detecting pixels. 
Phase zero corresponds to the timing of the radio's main pulse. 
A pulse shape similar to that observed in previous studies was obtained, characterized by an asymmetric main pulse preceding the radio main peak and interpulse.
The peak phase of the optical pulse is around $304\pm 35\,\mu$s considering the statistical error. 
As shown in Figure \ref{fig:lag}, this value is also consistent with previous studies \citep[e.g,][]{oo+06,Oosterbroek+08,Slowikowska+09}, and marginally consistent with \citet{str+13}.

\begin{figure}
    \centering
    \includegraphics[width=\linewidth]{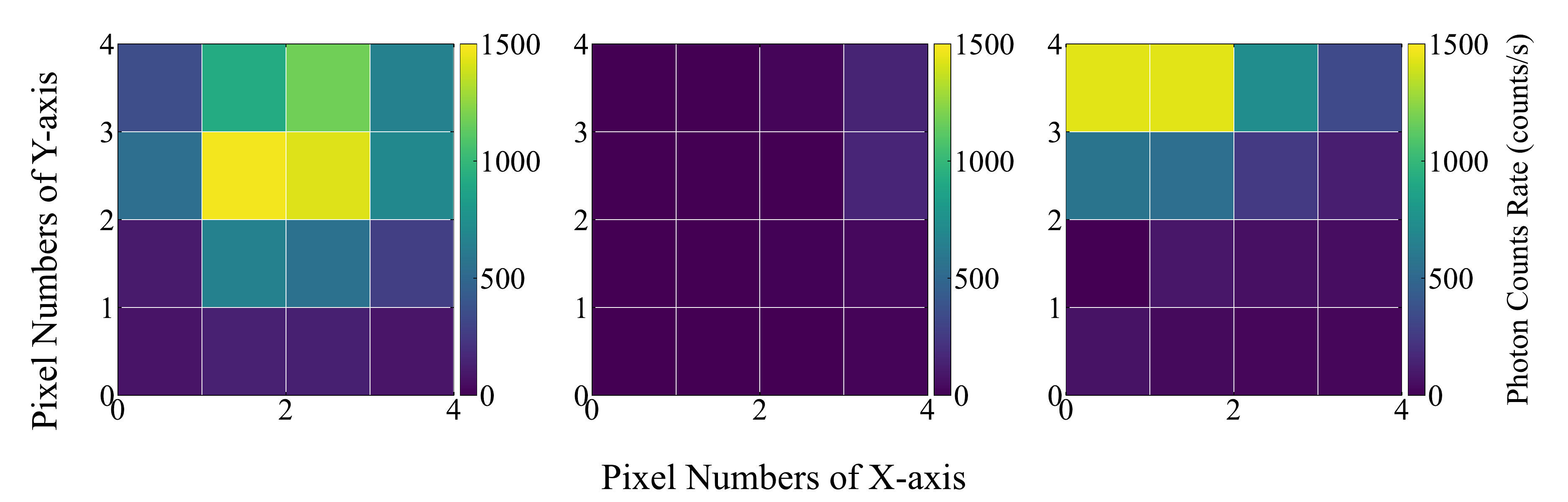}
    \caption{Optical image of the Crab pulsar obtained by the IMONY prototype.
    The dark subtraction and the flat correction were applied, and the exposure time was 3 minutes for each.
    (Left) On-pulse image. (Middle) Off-pulse image. (Right) On-pulse image with slightly mispointed.
    {Alt text: Three panels showing the count maps for all the $4\times 4$ pixels.}
    }
    
    \label{fig:optmaps}
\end{figure}

We reconstructed the phase-resolved images generated using only the counts of the pulse components.
Figure \ref{fig:optmaps} (left) and (center) show the on- and off-pulse images of the Crab pulsar, respectively, extracted from the same run data.
The spot size in the on-pulse image is consistent with a point source image with a typical seeing of 2--3",
and in the off-pulse range, no significant emission from the Crab pulsar was observed as expected.

\section{Discussion}

The IMONY prototype, 
which had been upgraded to an FPGA-based data acquisition system, 
was successfully installed on the Kanata telescope with the detection of the Crab pulsar period. 
The absolute time stamping capability is also confirmed by the fact that the Crab pulse peak timing was consistent with previous studies.
This work marked the first photon-counting observation conducted with the Kanata telescope.
Furthermore, we have demonstrated that IMONY worked as an imaging device.
In contrast, as noted in our previous paper \citep{naka+21},
the smaller number of pixels and narrow field of view still present a challenge.
Even during the observation of this work, 
approximately 60 minutes after the start of the observation, 
the object's image was no longer within the field of view. 
For example, in Figure \ref{fig:opt2d} (bottom), 
it is clear that the main pulse component was gradually disappearing before \textit{turns} $< 1.6\times 10^6$.
We also calculated the light curve of the observed pulsed component of the Crab pulsar.
Then we found the count rate varies from $\sim$ 1000 down to 50 counts/s.
This is because the Crab was moving out of the FoV,
as was apparent in the image during this time.
Figure \ref{fig:optmaps} (right) shows such an example of the Crab on-pulse image where its center is focused with an offset.
We also note that even in Figure \ref{fig:optmaps} (left) interval, 
the tail of the PSF is not contained in the FoV,
since the spot size is comparable to the sensor size.
We should conclude it is not straightforward that the current version of IMONY could measure the absolute flux of the source precisely. 
We might only mention that the highest observed count rate of $\sim 1$ kcounts/s is consistent with roughly estimated photon rate with V band magnitude of the Crab pulsar, 
the typical airmass absorption at the Kanata site, and the typical photon-detection-efficiency of the IMONY sensor.

Regarding the absolute timing of the optical peak,
our results of the peak timing depends on the uncertainty of $t_\mathrm{JPL}$.
Successful simultaneous high-frequency radio observations should be crucial 
to derive a reliable constraint on the stability of the optical peak timing, as discussed in \citet{kar07}.

As described in Section 2, the data transfer rate of the system was limited by the 100BASE-T capability. 
During the Crab observation in this work, the raw data rate was typically less than 2\,Mbit/s,
which was small enough for data taking and transferring all the data without loss.
If we observe sources brighter than 10 or 9 magnitudes, 
we must upgrade the readout board to 1000BASE-T or more.
From another point of view, this system is not suitable for such too-bright sources with high signal rates
since each pixel becomes insensitive to incident photons for $\sim$\SI{100}{\ns} after every Geiger discharge.
That is, there is an upper limit of observable target flux in principle.

In the case of observations such as the GRPs of the Crab pulsar, 
where the observations must be continued while awaiting the occurrence of a phenomenon, 
it is necessary to implement continuous tracking over an extended period. 
It is anticipated that a more careful pointing analysis 
than previously conducted would enhance the tracking accuracy. 
However, this approach is both laborious and ineffective. 
One potential solution would be to increase the size of the sensor and expand the field of view. 
A 64-pixel sensor system has already been developed 
and is currently undergoing testing \citep{hashi+24}. 
Due to this context, we do not focus on the accuracy of the photometry in this prototype
since the fraction of the detected photons was not stable during the observation.
The imaging capability turned out to be important 
since we were able to identify the drop in the flux was due 
not to the source variation but to the telescope's mispointing.
We intend to commence regular operation of IMONY as soon as possible, 
utilizing the system as a new system for optical photon-counting astronomy.

\begin{ack}
The authors express their sincere gratitude to Dr. A. Hayato for her
contribution in designing the jig of the system.
We also thank the anonymous referee for careful reading 
and giving helpful comments to improve the manuscript.

\end{ack}

\section*{Funding}
This work was supported by JSPS KAKENHI Grant Numbers JP20H10940 and JP23H01194, Yamada Science Foundation, and NAOJ Research Coordination Committee (NAOJRCC-2101-0103).

\section*{Data availability} 
 The data underlying this article are available at the appropriate request to the authors.

\end{document}